# Coherent strong-coupling of terahertz magnons and phonons in a Van der Waals antiferromagnetic insulator

**Authors:** Qi Zhang[1,2†*], Mykhaylo Ozerov[3†*], Emil Vinas Boström[4†], Jun Cui[1], Nishchay Suri[5], Qianni Jiang[2], Chong Wang[5], Fangliang Wu[1], Kyle Hwangbo[2], Jiun-Haw Chu[2], Di Xiao[5], Angel Rubio[4,6], Xiaodong Xu[2,7*]

[1]Department of Physics and National Laboratory of Solid State Microstructures, Nanjing University, Nanjing, 210093, China
[2]Department of Physics, University of Washington, Seattle, Washington 98195, USA
[3]National High Magnetic Field Laboratory, Florida State University, Tallahassee, Florida 32310, USA
[4]Max Planck Institute for the Structure and Dynamics of Matter, Luruper Chaussee 149, 22761 Hamburg, Germany
[5]Department of Physics, Carnegie Mellon University, Pittsburgh, Pennsylvania 15213, USA
[6]Center for Computational Quantum Physics, The Flatiron Institute, New York, NY 10010, USA
[7]Department of Materials Science and Engineering, University of Washington, Seattle, Washington 98195, USA

[†]These authors contributed equally to this work.

[*]Correspondence to: zhangqi@nju.edu.cn, ozerov@magnet.fsu.edu, xuxd@uw.edu

**Abstract:** Emergent cooperative motions of individual degrees of freedom, i.e. collective excitations, govern the low-energy response of system ground states under external stimulations, and play essential roles for understanding many-body phenomena in low-dimensional materials[1-3]. The hybridization of distinct collective modes provides a route towards coherent manipulation of coupled degrees of freedom and quantum phases[3,4]. In magnets, strong-coupling between collective spin and lattice excitations, i.e., magnons and phonons, can lead to coherent quasi-particle magnon polarons[5-11]. Here, we report the direct observation of a series of terahertz magnon polarons in a layered zigzag antiferromagnet FePS$_3$ via far-infrared (FIR) transmission measurements. The characteristic avoided-crossing behavior is clearly seen as the magnon-phonon detuning is continuously changed via Zeeman shift of the magnon mode. The coupling strength *g* is giant, achieving 120 GHz (0.5 meV), the largest value reported so far. Such a strong coupling leads to large ratio of *g* to the resonance frequency (g/ω) of 4.5%, and a value of 29 in cooperativity ($g^2/\gamma_{ph}\gamma_{mag}$). Experimental results are well reproduced by first-principle calculations, where the strong coupling is identified to arise from phonon-modulated anisotropic magnetic interactions due to spin-orbit coupling. These findings establish FePS$_3$ as an ideal testbed for exploring hybridization-induced topological magnonics in two-dimensions and the coherent control of spin and lattice degrees of freedom in the terahertz regime.

Magnons and phonons are the quanta of collective spin and lattice vibration, respectively. Besides their similarities in bosonic statistics and energy scales, the acoustic branches of magnons and phonons are the Goldstone modes of the system, with the spontaneous breaking of the spin rotational symmetry for the former and translational symmetry for the latter. The hybridization of magnons and phonons represents the coherent and dynamical coupling between the spin and lattice degrees of freedom. In the strong coupling regime, a new type of quasiparticles known as magnon polarons (MP) is formed. As shown in Fig. 1a, the coherent hybridization leads to the avoided crossing of the upper (UMP) and lower (LMP) magnon-polaron bands, whose energy difference is determined by the coupling strength $g$, in analogy to the vacuum Rabi splitting of light-matter interactions. The magnon polaron has recently attracted intense research attensions[5-10,12,13]. Theoretical studies suggested such hybridization enable Berry curvatures in the avoid-crossing regions of the magnon bands, and lead to nontrivial magnonic topology[5-7,9,10], as well as magnon mediated thermal Hall effects in two-dimensional magnets[9].

Besides intense theoretical investigations, signatures of magnon-phonon interaction have been identified via spin Seebeck measurements in ferromagnetic YIG[12], antiferromagnetic $Cr_2O_3$[13], and via inelastic neutron scattering in $CuCrO_2$[14] and $CuTeO_6$[15]. However, realization of magnon-phonon interactions in the strongly coupled regime is challenging, since it requires the coupling strength $g$ to be larger than the decay rate of individual modes. Hybrid structures, such as cavities[16,17] and surface patterning[18] have been utilized to create strongly coupled magnons and phonons. The avoided-crossing behavior, illustrated in Fig. 1a, is rarely seen in intrinsic materials[19]. In addition, previous studies mainly focus on magnon-acoustic phonon coupling via the Kittel-type magnetoelastic coupling with moderate coupling strength typically less than μeV[20] (or GHz regime). The hybridization of antiferromagnetic (AFM) magnons and optical phonons in the terahertz regime remains largely unexplored[21].

Recently, the emergence of van der Waals materials offer a new platform for studying and engineering coherent coupling between distinct degrees of freedom[3]. Examples include exciton polaritons up to room temperature[22,23], highly tunable phonon polaritons[24], and Dirac plasmon polariton[25,26]. However, the coupling between spin waves and collective excitations of other degrees of freedom remains to be explored. In this work, we report the observation of coherent strong-coupling between a series of magnons and phonons in the THz regime in a van der Waals antiferromagnetic insulator $FePS_3$. Remarkably, a giant coupling strength of 0.5 meV (~120 GHz) is achieved, 5 times larger than the MP linewidth. This giant coupling leads to the observation of clear avoided crossing of magnon and phonon branches, as well as the brightening of the IR-inactive Raman modes. These observations imply the long lifetime of THz magnons in $FePS_3$. Our first principle calculations well reproduce the observed MP dispersion curve, revealing its origin from the optical phonon-modulated anisotropic magnetic interactions due to spin-orbit coupling. At high magnetic field, we further observed a hysteretic spin-flip transition with the emergence of two ferromagnetic (FM) magnon branches, which yields a precise measurement of single ion anisotropy and interlayer exchange interactions.

The antiferromagnetic insulator of interest, $FePS_3$, belongs to a class of transition metal phosphorous trichalcogenides ($MPX_3$, $M$ = Fe, Mn, Ni and $X$ = S, Se). Magnetic moments mainly come from the Fe atoms, arranged in a honeycomb spin lattice structure[27-32]. It has a zigzag AFM ground state with strong magnetic anisotropy, and is considered as a 2D Ising spin system[30,31]. We performed far-infrared magnetospectroscopy (FIRMS) measurements on single crystal $FePS_3$ flakes of ~ 50 μm in thickness (methods). The magnetic field is applied out-of-plane, parallel with the spin directions (Faraday geometry). The measurement was performed at 4.2 K, well below the antiferromagnetic transition temperature ($T_N \sim 117K$). Normalized FIR transmission spectra as a function of magnetic fields are presented in Fig. 1b (See Supplementary Figure S1 for raw data). At zero field, a prominence mode at 3.67 THz (122 cm$^{-1}$) is the doubly degenerated acoustic AFM magnons, which is consistent with previous Raman study[33]. The optical AFM magnons have also been observed at 9.6 THz (320 cm$^{-1}$) (See Supplementary Figure S2).

When an external magnetic field was applied, the degeneracy of the AFM magnons with opposite angular momenta was lifted, and the AFM magnon modes exhibit Zeeman splitting with a $g$-factor of 2.1. As the field increasing, the low-energy AFM magnon branch approaches a nearby phonon mode (Ph3) located at 3.25 THz. This mode is a zone-folded optical phonon at the $\Gamma$-point with its atomic motions shown in Fig. 1a, bottom panel. In the paramagnetic phase of FePS$_3$, it corresponds to an acoustic mode at M-point (see Supplementary Figs. S3 and S4 for temperature and polarization resolved study of the Raman modes).[30] In the case of no interaction, these two modes will cross around 13 T (zero-detuning). Experimentally, however, a clear avoided-crossing behavior is observed, which is the signature of coherent magnon-phonon interaction in the strong-coupling regime.

The top-panel of Fig. 1b shows the FIR spectrum at 13 T, which is close to the zero detuning. The separation between UMP and LMP peaks is about 170 GHz. The magnon-phonon coupling strength $g$ was further extracted to be 85 GHz (see supplement Text S1), which is much larger than the linewidth of both the UMP (26 GHz) and LMP (24 GHz). The small dip next to the LMP is due to the coupling between the magnon and a phonon at 3.15 THz with a coupling strength (~10 GHz) slightly smaller than the MP linewidth. Figure 1c shows the extracted MP linewidth as a function of detuning. As the detuning is decreased, the linewidths of UMP and LMP modes approach each other, and finally cross at the zero detuning, evidencing the coherent hybridization of magnons and phonons.

Multiple magnon polarons are also observed in high-field FIR transmission measurements up to 35 T, as shown in Fig. 2a. As the magnetic field increases above 20T, the low-energy branch of the AFM magnon exhibits a series of avoided crossing with nearby phonon modes. Similar to Ph3 mode (3.25 THz), Ph1 (2.63 THz) and Ph2 (2.83 THz) are also Brillouin zone-folded phonons (from M point to $\Gamma$ point) due to the formation of the zigzag spin order. It is revealed by the temperature dependent Raman scattering and supported by first principle calculations (Supplement Fig. S3). The atomic motions of Ph1, Ph2 and Ph3 phonons are illustrated in Supplement Fig.S5.

The avoided crossing and the MP branches are fitted with a theoretical model discussed below (Eq.2) for coupled magnon/phonon modes (also see Supplement Text S1). As shown in Fig. 2b, experimental magnon/phonon peak positions are well described by the fitting. The fitting lines are color-coded, which indicates the percentage of magnon/phonon components in each hybridized MP branch. The greenish regions represent nearly 50%/50% hybridization of magnons and phonons. The composition of MP branch is shown in Fig.2c, where MP is decomposed into uncoupled magnons and phonons with their percentage given by the normalized amplitude square of each mode (see Supplement Text S1). Uncoupled magnon/phonon frequencies and the coupling strength $g$ for each magnon-phonon pair are extracted and listed in Table 1. A $g$ value of 120 GHz was obtained between the magnon and Ph1 phonon, which is the largest value reported so far. The ratio of coupling strength and resonance frequencies ($g/\omega$) reaches 4.5%. The cooperativity, defined as $g^2/\gamma_{ph}\gamma_{mag}$, is a dimensionless merit of quantifying the efficiency of energy circulation between the coupled modes. It reaches a value of 29 in our measurements, which is remarkable for coherent magnon-phonon coupling systems.

Now we consider the symmetry requirements for the magnon-phonon strong coupling. Monolayer FePS$_3$ has the point group of $D_{3d}$. The symmetry reduces to the space group *C2/m* due to the formation of the zigzag spin order and the monoclinic stacking. The *C2/m* point group has four irreducible representations, $\Gamma_1$ to $\Gamma_4$. The IR excitation of magnons in FePS$_3$ is via magnetic dipole interaction. Its transition matrix element possesses the irreducible representation $\Gamma_1$ with even parity (see Supplement Text S2). In order to have the avoided crossing and mode hybridization, the phonons involved in the magnon-phonon strong coupling must belong to the same irreducible representation with even parity, namely, they are Raman-active modes. This understanding is supported by our Raman measurements, where those phonon modes indeed exhibit strong Raman signals (Supplementary Fig. S3). In general, Raman-active modes with even parity are IR forbidden. The reasons for our observation of these Raman-active phonons in FIR transmission spectra are two fold. First, the relatively large sample thickness and high-order effects (e.g., electric-quadrapole transitions) make IR-inactive modes visible in IR transmission spectra. More importantly, those phonon modes gain oscillator strength due to the hybridization with IR-active magnons. This brightening effect can be seen clearly in the strong-coupling region in our spectra.

A microscopic model of MP can be derived by considering the spatial dependence of the magnetic interactions. The low-energy excitations of FePS$_3$ are described by the following Hamiltonian[34-37],

$$H = \sum_{<ij>} \mathbf{S}_i \cdot (\mathbf{J}_{ij} \cdot \mathbf{S}_j) - \Delta \sum_i (S_i^z)^2 + \sum_{ij} \left[ \frac{\hat{p}_i^2}{2M} + \frac{k_{ij}}{2} \hat{x}_{ij}^2 \right] \quad (1)$$

Where the matrix $\mathbf{J}_{ij}$ encodes the coupling of spins $\mathbf{S}_i$ and $\mathbf{S}_j$, and $\Delta$ is the single ion anisotropy, $\hat{p}_i$ is the momentum of an ion of mass $M$, $\hat{x}_i$ is the ionic displacement, $\hat{x}_{ij} = \hat{x}_j - \hat{x}_i$, and $k_{ij}$ is the elastic tensor. The magnon-phonon interaction enters through the dependence of the interactions on the ion displacement, i.e., $\mathbf{J}_{ij} = \mathbf{J}_{ij}(\hat{x}_{ij})$.

Due to the negligibly small photon momenta, the observed strong-coupling locates at the $\Gamma$ point ($k = 0$). This rules out the most studied Kittel-type magnon-phonon interactions, which vanishes at $k = 0$[20]. The coherent magnon-phonon hybridization requires quadratic coupling terms, e.g., $(a^\dagger + a)(b^\dagger + b)$, where $a$ ($b$) is the annihilation operator of magnons (phonons). However, for diagonal Heisenberg exchange, i.e., $J_{xx}$, $J_{yy}$ and $J_{zz}$, expanding $\mathbf{J}_{ij}(\hat{x}_{ij})$ in terms of $\hat{x}_{ij}$ only gives an $a^\dagger a(b^\dagger + b)$ type coupling, which renormalizes the magnon frequency rather than provides avoided crossing. In contrast, the phonon-induced spatial modulation of anisotropic (off-diagonal) interactions, e.g., $J_{xz}$ and $J_{yz}$, provide quadratic coupling terms. Including such terms and transforming the Hamiltonian to the magnon and phonon basis, the MP excitations at the zone center are described by the Hamiltonian,

$$H = \sum_{i=1,2} \omega_{mag,i} \hbar a_i^\dagger a_i + \sum_\lambda \omega_{ph,\lambda} \hbar b_\lambda^\dagger b_\lambda - \sum_\lambda (b_\lambda^\dagger + b_\lambda)[g_{\lambda 1}(a_1^\dagger + a_1) + g_{\lambda 2}(a_2^\dagger + a_2)] \quad (2)$$

In this equation, the first (second) term describes for the magnon (phonon) energy. The third term represents the coherent magnon-phonon coupling with the coupling strength $g_{\lambda i}$.

To describe the MP formation from the first principles, we combine the above microscopic model with parameters from *ab-initio* simulations (see Methods). Calculated MP branches as a function of magnetic field are shown in Fig. 3b. Avoided crossings are well reproduced. The calculated magnetic interactions of the equilibrium system (schematically shown in Fig.3a, values are listed in the supplement Table S1), are in good agreement with recent neutron scattering data[32,38], and give two degenerate magnon modes at B = 0 T at 3.71 THz. Similarly, the phonon frequencies agree well with previous theoretical studies[30,39], and we find three relevant zone-folded modes in the region of interest at 2.70 THz, 2.82 THz and 3.26 THz. The corresponding coupling strength with magnons are 93, 61 and 83 GHz respectively. Remarkably, these first-principle values match well with experimental ones, as listed in Table 1. Such agreement further justifies our microscopic model, and supports the fact that the magnon-phonon coupling in FePS$_3$ originates from the phonon-modulated anisotropic magnetic interactions. We note that the zigzag magnetic order is crucial for the formation of MP in FePS$_3$. In fact, for the structurally similar but Néel-ordered MnPS$_3$, our MP model and first principles calculations exhibit vanishing magnon-phonon coupling. This is due to the additional symmetries of the Néel state leading to a cancellation of the coupling to the neighboring spins. Such symmetry is broken in the zigzag state (see Supplement Text S4).

On the microscopic level, anisotropic magnetic interactions can arise from magnetic dipole-dipole interactions and/or from spin-orbit coupling (SOC) mediated exchange or coupling to the crystal field. However, since the magnetic dipole terms have the form $(S_i \cdot R_{ij})(S_j \cdot R_{ij})$, where $R_{ij}$ is the equilibrium ionic distance in the *xy*-plane between ion *i* and *j*, thus this interaction cannot give rise to $J_{xz}$ and $J_{yz}$. Indeed, by performing calculations with and without SOC, we find that both the magnetic anisotropies $J_{xz}$ and $J_{yz}$, and the magnon-phonon coupling, vanishes when SOC is neglected. This analysis and our first principles calculations demonstrate that the strong magnon-phonon coupling in FePS$_3$ comes from SOC, via phonon-modulation of either the crystal field or anisotropic exchange interactions. Both the zigzag spin order and SOC are thus crucial for magnon polaron to form in this material.

In Fig. 2a, a sudden change in the FIR spectrum appears at 35 T. This is a signature of a first-order AFM-to-FM spin-flip transition in FePS$_3$. Two new modes appear at 3.77 THz and 3.88 THz (features near 16 meV) corresponding to two FM magnons. Those FM magnons can be further confirmed as they survived during field down-sweep until 25 T as presented in Fig.4a. Such hysteretic behavior with magnetic field sweeping direction[40], along with the co-existence of AFM and FM modes in the spectra indicate the first-order nature of this AFM-FM magnetic transition and the formation of domains. In contrast to the usual spin-flop type transitions, where the FM magnon emerges from zero frequency, the observed FM magnons appears suddenly at finite frequencies. This is the signature of a spin-flip transition, which happens in AFM with strong magnetic crystalline anisotropy ($\Delta > J$)[41]. The separation between the two FM magnons originates from the interlayer exchange coupling $J'$. By applying linear spin wave theory (Supplementary Text S3), we obtain the FM magnon frequencies, $\omega_F = 2S(\Delta \pm 6J') + g\mu_B H$, where *S* is the spin angular momentum and $g\mu_B H$ is the Zeeman energy. Hence, through our measurements, the $\Delta$ and $J'$ are accurately determined to be $2.92 \pm 0.02$ meV and $-9.4 \pm 0.1$ μeV, respectively.

The value of $J'$ is more than 140 times weaker than the intralayer exchange interactions (e.g., $J_1$), making FePS$_3$ an idea 2D spin system even in the bulk form.

Our work revealed a series of coherent strongly coupled magnons and zone-center optical phonons with unprecedent coupling strength. The microscopic origin of such coupling is identified to be the phonon-modulated anisotropic magnetic interactions. These findings open routes toward effective coherent manipulation of magnon Berry curvatures and topology via magnon-phonon interactions. Combined with the existing toolbox of vdWs materials engineering, e.g., electrostatic gating, strain, heterostructures and twisting, 2D antiferromagnetic $M$PS$_3$ system is established as a promising platform for exploring correlated phenomena with coherent strongly coupled degrees of freedom, as well as spin-lattice engineered AFM spintronics in the terahertz regime.

**Acknowledgements:** Authors thank Dr. Yuan Wan for valuable discussions. This project is mainly supported by the Department of Energy, Basic Energy Sciences, Materials Sciences and Engineering Division (DE-SC0012509). QZ's work is partially supported from MOST of China (Grant No. 2020YFA0309200) and Fundamental Research Funds for the Central Universities (0204-14380184). The FIR measurement was performed at the NHMFL which is supported by the NSF (Grant number DMR1644779) and the State of Florida, US. First principle work was supported by the European Research Council (ERC-2015-AdG694097), the Cluster of Excellence 'Advanced Imaging of Matter' (AIM), Grupos Consolidados (IT1249-19) and SFB925 "Light induced dynamics and control of correlated quantum systems". AR acknowledges support from the Max Planck-New York City Center for Non-Equilibrium Quantum Phenomena. The Flatiron Institute is a division of the Simons Foundation. Bulk crystal growth is supported by NSF MRSEC DMR-1719797 and the Gordon and Betty Moore Foundation's EPiQS Initiative, Grant GBMF6759 to JHC.

**Author contributions:** QZ and XX conceived the experiment. MO performed the FIR high-field measurement. JC, FW, QZ and KH carried out Raman measurements. QZ analyzed the data with the help from JC. EVB and AR conduct first-principle calculations. NS, CW and DX performed symmetry analysis and spin-wave calculations. QJ and JHC synthesized the crystals. QZ and XX wrote the paper with input from all authors. All authors discussed the results.

**Competing Interests:** The authors declare no competing financial interests.

**Data Availability:** Source data for the figures in the main text are provided with the paper. All other data that support the plots within this paper and other findings of this study are available from the corresponding author upon reasonable request.

**Methods:**

**Crystal growth and sample fabrication**: Single crystals of $FePS_3$ were synthesized by chemical vapor transport (CVT) method using iodine as the transport agent. Stoichiometric amounts of Iron powder (99.998%), phosphorous powder (98.9%) and sulfur pieces (99.9995%) were mixed with iodine (1 mg/cc) and sealed in quartz tubes (10 cm in length) under high vacuum. The tubes were placed in a horizontal one-zone tube furnace with the charge near the center of the furnace. Sizeable crystals ($10.0 \times 10.0 \times 0.5$ mm$^3$) were obtained after gradually heating the precursor up to 750 °C, dwelling for a week and cooling down to room temperature.

**Far-Infrared Magnetospectroscopy (FIRMS) measurements:** FIRMS measurements were performed at the National High Magnetic Field Laboratory in Tallahassee, FL using combinations of a) Bruker Vertex 80v FT-IR spectrometer with 17 T superconducting magnet and b) Bruker IFS66 FT-IR spectrometer with 35 T resistive magnet. The far-IR radiation was propagated inside an evacuated optical beam line from the spectrometers to the top of the magnets. Thereon, the light pipe guided the radiation down to the sample located at the center of the magnet and cooled down to 5 K by low-pressure helium exchange gas. The transmission was detected by a 4.2K Si bolometer placed in the subtle fringe field behind the sample in order to minimize loss of intensity. The Faraday geometry (the incident radiation was parallel to the applied field) was employed in both experimental set-ups. All FT-IR spectra were recorded in the frequency range of 10-720 cm$^{-1}$ with a resolution of 0.3 cm$^{-1}$. To improve the signal-to-noise ratio, each FT-IR measurement was repeated two times at each field step, then averaged. In order to unveil field-dependent excitations from those that are field-independent, spectra at each magnetic field step were divided by the average of all spectra, resulting in FIRMS spectra with clear "magnetic" spectral features above a more-or-less flat baseline and successful suppression of field-independent background in the transmittance.

**Raman measurements**: A HeNe laser (632.8 nm) was used to excite the $FePS_3$ sample, which was placed in a closed cycle cryostat with temperature range from 5 K to 300 K with a superconducting magnet up to 9 T. Raman measurements were polarization-resolved and collected by a spectrometer with a liquid nitrogen cooled CCD camera. An out-of-plane magnetic field up to 9 T was applied (Faraday geometry).

**First-principle calculations**: To obtain the magnon-phonon coupling of bulk $FePS_3$, we performed first principles simulations with the ABINIT electronic structure code[42-44]. We used the local density approximation with projector augmented wave (PAW) pseudopotentials, a plane wave cut-off of 30 Ha and 60 Ha respectively for the plane wave and PAW part, and included an empirical Hubbard $U$ of 2.7 eV self-consistently determined in OCTOPUS via the ACBN0 functional[45,46]. A $\Gamma$-centered Monkhorst-Pack grid with dimensions 8*4*5 was used to sample the Brillouin zone.

The ground state was found to have zigzag antiferromagnetic order with spins aligned along the z-axis. The DFT wave functions were mapped onto effective localized orbitals via the WANNIER90 code, and the spin parameters were calculated via the magnetic force theorem and the Green's function methods as implemented in the Python package TB2J[47-49]. The resulting spin parameters are in good agreement with recent neutron scattering[32,38]. Phonon frequencies and eigenvectors were calculated with ABINIT for a ferromagnetic interlayer coupling, after relaxing the atomic positions and stresses to below $10^{-6}$ Ha/Bohr, and are in good agreement with previous DFT calculations and Raman scattering data[30,39].

Table 1

| | ω (THz) Undressed | γ (GHz) at 0T | g (GHz) Exp. | Cooperativity $\frac{g^2}{\gamma_{mag}\gamma_{ph}}$ | g (GHz) Cal. |
|---|---|---|---|---|---|
| **AFM magnon** | 3.64 | 35 | — | — | — |
| **Ph1 phonon** | 2.64 | 14 | 120 | 29 | 93 |
| **Ph2 phonon** | 2.83 | 16 | 65 | 7.5 | 61 |
| **Ph3 phonon** | 3.27 | 16 | 85 | 15 | 83 |

**Table 1 | Magnon-phonon strong-coupling parameters of FePS$_3$.**

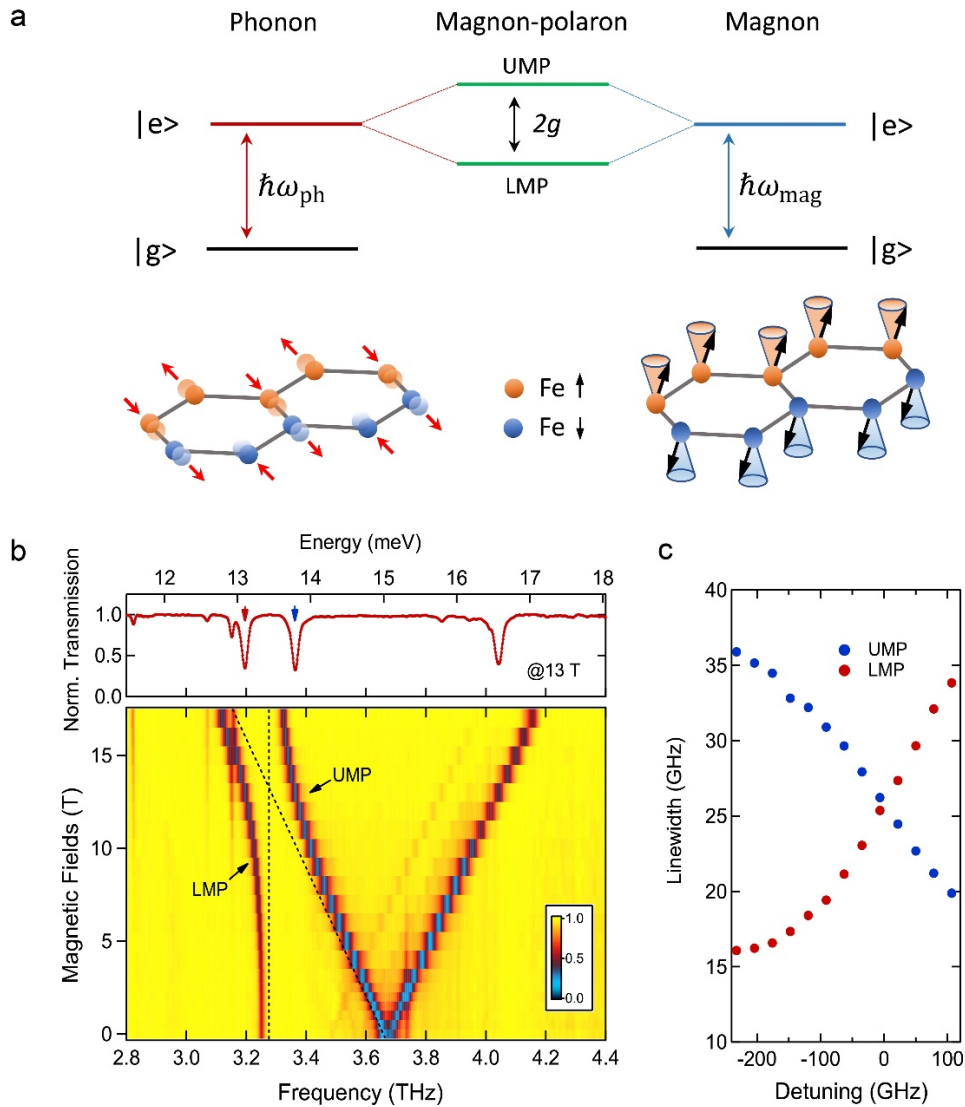

**Figure 1 | Strongly-coupled terahertz antiferromagnetic magnons and phonons in FePS$_3$. a,** Up panel: schematics of coherent strong-coupling between magnons and phonons, which leads to the formation of upper and lower magnon-polaron bands (UMP and LMP), separated by the coupling strength 2*g*. Bottom panel: schematic atomic motions of the related phonon mode (Ph3) at 3.25 THz. The zigzag spin chain is color coded. The spin motion of the AFM magnon (3.67 THz at 0T) is also shown. **b,** Normalized magneto-FIR transmission spectra of FePS$_3$ with out-of-plane magnetic fields up to 17 Tesla. Clear avoided crossing between the Ph3 phonon and the low-energy branch of the Zeeman-split AFM magnon is observed. Black dotted-lines indicate the uncoupled magnon and phonon modes. The top panel shows the transmission spectrum at 13T, which is close to the zero detuning between the uncoupled magnon and phonon modes. The UMP (LMP) peak is indicated by blue (red) arrows. **c,** The linewidths of upper (blue) and lower (red) magnon polarons as a function of detuning, which crosses near the zero-detuning point.

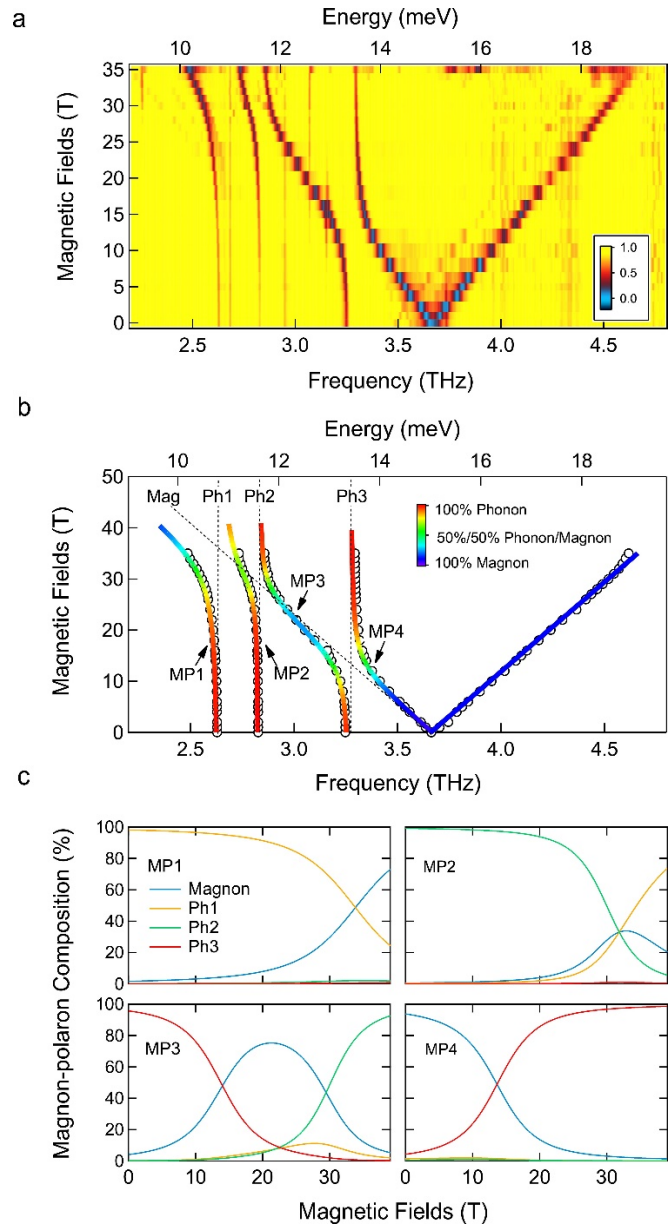

**Figure 2 | Observation of multiple magnon polarons up to 35T. a,** Normalized magneto-FIR transmission spectra of FePS$_3$ measured with magnetic field sweeping from 0 to 35T. As magnetic fields increase, the AFM magnon exhibit strong-coupling with Ph3, Ph2 and Ph1 phonons in sequence. At 35T, new ferromagnetic (FM) magnon modes emerged around 3.8 THz. **b**, Fitted magnon-polaron branches as a function of magnetic field. Fitting lines are color coded by their magnon/phonon composition. As indicated by the color bar, the blue (red) color indicates dominated magnon (phonon) components, while the greenish color shows the coherently coupled magnons and phonons with nearly equal amplitude. Experimental peak positions are shown in open circles. Uncoupled magnon/phonon modes are shown by the dotted lines. Magnon-polaron branches (MP1 to MP4) are indicated by black arrows. **c**, The composition of magnon polarons as a function of magnetic fields. Each magnon-polaron state is decomposed into uncoupled magnon/phonon modes. (see Supplement Text S1).

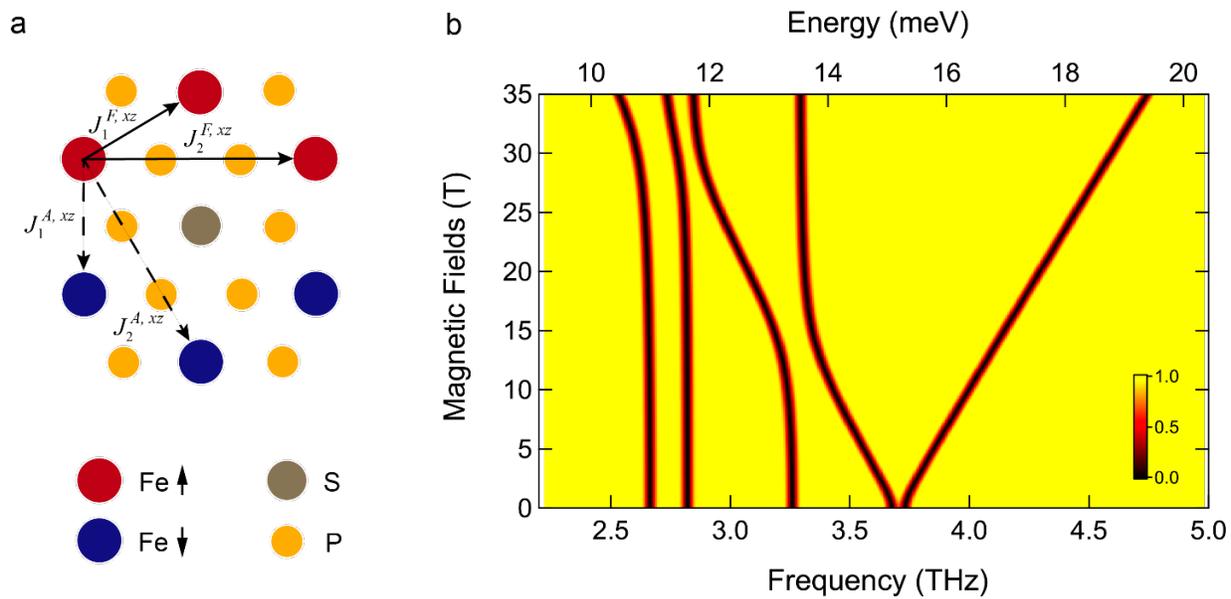

**Figure 3 | First-principle calculations of coherent strong couplings between magnon and phonons in FePS$_3$. a,** Schematic diagram of the nearest and next nearest off-diagonal exchange interactions. **b**, Calculated magnon dispersion as a function of magnetic fields. The avoided crossings between the magnon and Ph1 to Ph3 phonons are well reproduced with calculated coupling strength of 93, 61 and 83 GHz, respectively.

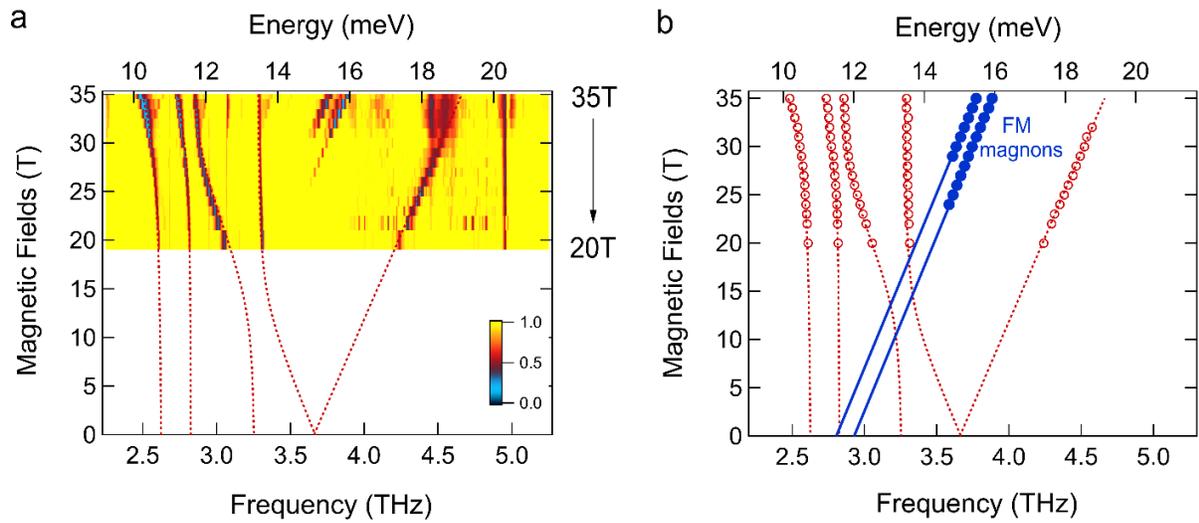

**Figure 4 |Spin-flip transitions of FePS$_3$ at high magnetic fields. a,** FIR spectra of FePS$_3$ as magnetic field decreasing from 35 T to 20 T. Two FM magnons survive until 25 T. The separation of the two FM modes originates from the interlayer exchange coupling. **b,** Peak positions of FM magnons and their linear spin wave fittings (blue solid lines). Single ion anisotropy $\Delta$, and interlayer exchange interaction $J'$ can be accurately determined. Red dotted lines are fitted magnon-polaron branches, the same as the ones shown in Fig.2b.